\documentclass[12pt]{article}

\usepackage{amssymb,amsfonts,amsmath,epsfig}
\usepackage{graphicx} 
\usepackage{indentfirst}
\usepackage{bbm}
\usepackage{multirow}

\parskip=\medskipamount

\newcommand{\sect}[1]{\setcounter{equation}{0}\section{#1}}

\newcommand{\be}{\begin{equation}}
\newcommand{\ee}{\end{equation}}
\newcommand{\bea}{\begin{eqnarray}}
\newcommand{\eea}{\end{eqnarray}}

\def\double #1{#1{\hbox{\kern-2pt $#1$}}}

\def\gkk{g^{\rm (1)}_{\rm KK}}

\begin{document}

\begin{titlepage}

\begin{flushright}
arXiv:12mm.xxxx [hep-ph]
\\
{\bf OCHA-PP-312}
\end{flushright}
\vspace{2mm}

\begin{center}
{\Large \bf  LHC sensitivity to Kalzua-Klein gluon \\[3mm] in two 
 $b$-jets decay channel} 
\end{center} 

\begin{center}
{\bf
Masato Arai$^\dagger$$^\sharp$\footnote{masato.arai@fukushima-nct.ac.jp},
Gi-Chol Cho$^\ddagger$\footnote{cho.gichol@ocha.ac.jp},
Karel Smolek$^\sharp$\footnote{karel.smolek@utef.cvut.cz},
Kyoko Yoneyama$^\ast$$^\ddagger$\footnote{yoneyama@hep.phys.ocha.ac.jp}
} \\
\vspace{3mm}
{\it $^\dagger$Fukushima National College of Technology,
Fukushima, 970-8034, Japan} \\
{\it $^\ddagger$Department of Physics, Ochanomizu University, Tokyo 112-8610, Japan}\\
{\it $^\sharp$Institute of Experimental and Applied Physics,
Czech Technical University in Prague, 
Horsk\' a 3a/22, 128 00 Prague 2, Czech Republic}\\
{\it $^\ast$Department of Physics, Bergische Universit\"at Wuppertal
Gaussstr. 20, D-42119 Wuppertal, Germany}
\\

\vspace{2mm}

\end{center}
\vspace{5mm}

\begin{abstract}
\baselineskip=14pt
\noindent

We examine a possibility to discover a Kaluza-Klein (KK) excitation of 
 gluon in a warped extra dimension model at the Large Hadron Collider  
focusing on a decay channel of the KK gluon into a $b$-quark pair. 
It is known that, in a certain extension of the warped extra dimension 
 model, the third generation quarks could strongly couple to the 
KK gluon owing to appropriate bulk fermion mass parameters. 
Taking account of kinematical cuts to reduce background events,  
we show the model parameter space which leads to a significance 
larger than 5$\sigma$ with the integrated luminosity of 
10~(100) ${\rm fb}^{-1}$. 
\end{abstract}
\vspace{1cm}

\vfill
\end{titlepage}

\newpage
\setcounter{page}{1}
\renewcommand{\thefootnote}{\arabic{footnote}}
\setcounter{footnote}{0}

\tableofcontents{}
\vspace{1cm}
\bigskip\hrule

\sect{Introduction}
Although the Standard Model (SM) of particle physics has shown a good 
agreement with almost all data of high energy experiments, we expect new 
physics beyond the SM from some theoretical motivations such as a gauge 
hierarchy problem.  
A warped extra-dimension model proposed by Randall and Sundrum 
(RS)~\cite{RS} is one of the promising candidates to explain the 
hierarchy between the Fermi scale and the Planck scale. 
In the RS model, there are two 3-branes which are located in different  
positions in the fifth dimension. 
One of the 3-branes is called a ``visible brane'' in which the SM 
particles are confined while the other is called a ``hidden brane''.  
A graviton is allowed to propagate between two branes. 
With this set-up, the mass scale of a Higgs boson can be electroweak 
scale naturally when the fifth dimension is warped appropriately and the
gauge hierarchy problem is understood without suffering from a 
fine-tuning problem like the SM. 
Although it is sufficient to explain the gauge hierarchy problem when 
only the graviton propagates into the extra dimension, an extension of
the RS model, where some of the SM particles also propagate 
into the bulk, has been studied from a phenomenological point of view 
(for example, see \cite{Davoudiasl:2000wi}). 
A generic consequence of such an extension of the RS model is that 
there are Kaluza-Klein (KK) excitations of the SM particles. 
It is, therefore, important to investigate possibilities of production
and decay of KK particles at the Large Hadron Collider 
(LHC) as a direct test of the model.  

The mass of KK particles are typically determined by the scale 
parameter $\Lambda_{\rm KK}$. 
It has been shown that the KK excitations of the electroweak gauge
bosons significantly constrained from the electroweak precision
measurements due to their large contributions to the oblique parameters. 
As a result, the scale of the KK mode $\Lambda_{\rm KK}$ is required to be 
$\Lambda_{\rm KK}> O(10^{2-3}{\ \rm TeV})$, 
which leads to an unwanted hierarchy between the electroweak scale 
$\Lambda_{\rm EW}\sim O(m_W)$ 
and $\Lambda_{\rm KK}$~\cite{Davoudiasl:2000wi,Davoudiasl:1999tf}.  
Such a constraint could be somewhat lowered to $O({\rm TeV})$ by 
introducing the custodial symmetry in the bulk \cite{Agashe:2003zs} 
and additional contributions  
from the bulk SM fermions~\cite{HePeRi}. 
Phenomenology of introducing the bulk custodial symmetry 
has been studied, {\it e.g.}, in ref.~\cite{Djouadi:2006rk}. 

Another serious constraint on the KK scale $\Lambda_{\rm KK}$ comes from  
processes mediated by flavor changing neutral currents (FCNC).  
For example, a naive estimation of contributions of KK gauge bosons
to a CP violating parameter $\epsilon_K$ in the $K^0$-$\overline{K^0}$ 
mixing tells us that the KK scale should be $\Lambda_{\rm KK}>20$ TeV. 
However, the bound on $\Lambda_{\rm KK}$ could be lowered to the scale 
which is allowed  
from the electroweak precision data arranging the fermion sector
of the model such as appropriate choice of the bulk fermion masses
or introducing some flavor symmetry 
(a useful and compact summary has been given in ref.~\cite{JuWe}). 
It should be noted that there is another study that the KK gauge boson
mass could be $O({\rm TeV})$ without introducing the custodial
symmetry on the bulk~\cite{Carmona:2011ib}.  
Thus we expect that the KK gauge bosons with a few TeV mass could be
produced at the LHC and it is worth examining signatures of these new
particles using various decay channels. 

In this paper, we study a possibility of observation of the first KK 
excitation of gluon ($\gkk$)  at the LHC using the decay channel to two
$b$-quarks, $\gkk \to b \bar{b}$. 
Searching for $\gkk$ at the LHC has been studied using a decay of $\gkk$ 
into a top-quark pair, $\gkk \to t \bar{t}$, {\it e.g.}, in refs.~\cite{AgBeKrPeVi, 
GuMaSr1, LiRaWa, JuWe}, since there is a naive expectation that the 
interaction of the right-handed top-quark $t_R$ with the KK gluon less  
affects the electroweak and flavor processes so that the large coupling 
of $t_R$ could be allowed phenomenologically. 
On the other hand, the $b$-quark couplings to the KK gluon 
has been chosen to be negligibly suppressed in the literatures 
due to the FCNC constraints. 
In our analysis, however, 
since the Yukawa sector of the RS model is still controversial ({\it e.g.}, 
ref.~\cite{Chang:2008vx}), we adopt the left- and  
right-handed $b$-quark couplings to the KK gluon as phenomenological 
parameters and  
study the possibility to observe the KK gluon through the two 
$b$-jets channel at the LHC. 

As will be shown later, a single production of $\gkk$ is possible 
only through pair-annihilation of quarks. 
Thus, its production in the $s$-channel is relatively suppressed 
when the couplings of light quarks to the KK gluon is small 
so that the experimental lower bound on $\gkk$ is 
of order 1 TeV in both Tevatron \cite{CDF} and the LHC \cite{CMS,ATLAS}. 
Note that these experimental bounds have been obtained using the decay 
channel $\gkk \to t\bar{t}$. 
The associate production of $\gkk$ decaying to $t\bar{t}$ has been
studied in ref.~\cite{GuMaSr2}. 

In our study, we focus the following process $pp \to \gkk \to b\bar{b}$. 
In general, it is hard to find a signal process using two
$b$-jets because of a huge QCD background. 
We find that, imposing efficient kinematical cuts, extraction of signal
events from the background is possible. 
For example, a significance could be larger than 5$\sigma$ for the
$\gkk$ mass up to $1.1~{\rm TeV}$ with the integrated luminosity of $10~{\rm
fb}^{-1}$ and $1.4~{\rm TeV}$ with $100~{\rm fb}^{-1}$. 

This paper is organized as follows. 
In the next section, we briefly review the RS model with bulk fermions
and gauge bosons. 
Our numerical results are shown in Sec. 3. 
Sec. 4 is devoted to the summary and discussion. 

\sect{Model setup}
We consider a five-dimensional space-time with a non-factorizable geometry \footnote{We follow the convention in ref. \cite{GhPo}.}
\begin{eqnarray}
 ds^2=e^{-2\sigma}\eta_{\mu\nu}dx^\mu dx^\nu-dy^2,\quad \mu,\nu=0,1,2,3, \label{metric}
\end{eqnarray}
where $\sigma=k|y|$, $\eta_{\mu\nu}={\rm diag}(-1,1,1,1)$, $y$ is the coordinate of the fifth dimension and 
$k$ determines the curvature of the $AdS_5$. The coordinate $y$ is compactified on 
an orbifold $S^1/{\bf Z}_2$ of a radius $r_c$, with $-\pi r_c\le y \le \pi r_c$. 
The orbifold fixed points at $y=0$ and
$y=\pi r_c$ are locations of two 3-branes, which are called the hidden brane and 
the visible brane, respectively. At the visible brane, 
the effective mass scale is given to be $M_P e^{-\pi kr_c}$, 
associated with the TeV scale provided $kr_c\simeq 12$. 
Note that $M_P$ is the four-dimensional Planck scale. 
Thus the gauge hierarchy problem is solved in this model. 
In our scenario, we assume that the SM Higgs is located at the visible 
brane while the other SM fields and the gravity are present in the 
five-dimensional bulk.  
We are interested in the case that the third generation of quarks 
couples to the KK gluons strongly. The relevant part of the model to our 
analysis is then written by an $SU(3)$ gauge field $A_M^a$ and a Dirac 
fermion $\Psi$ with the five-dimensional coordinates labelled by capital
Latin letters, $M=(\mu,y)$ and the adjoint index of the gauge group, 
$a$. 
The five-dimensional bulk action of the gauge field and fermion is 
given by 
\begin{eqnarray}
S_5=-\int d^4x \int dy \sqrt{-G}\left[-{1 \over 4}F_{MN}^aF^{MNa}+i\bar{\Psi}\gamma^M D_M\Psi+i m_\Psi \bar{\Psi}\Psi \right], \label{action}
\end{eqnarray}
where $G=\det{(G_{MN})}$, the gauge field strength is defined by $F_{MN}^a=\partial_MA_N^a-\partial_NA_M^a+if_{abc}A_M^bA_N^c$ with the structure constant $f^{abc}$ and $\gamma_M=(\gamma_\mu, \gamma_5)$ is defined in curved space as $\gamma_M=e^\alpha_M\gamma_\alpha$, where $e_M^\alpha$ is the funfbein and $\gamma_\alpha$ are the Dirac matrices in a flat space. Covariant derivative is written by $D_M=\partial_M+\Gamma_M+ig_5A_M$ where $\Gamma_M$ is the spin connection and $g_5$ is the 5-dimensional  gauge coupling constant. For the metric (\ref{metric}), the spin connection is given by $\Gamma_\mu={1 \over 2}\gamma_5\gamma_\mu {d\sigma \over dy}$ and $\Gamma_5=0$. The bulk fermion mass $m_{\Psi}$ is parameterized as
\begin{eqnarray}
  m_{\Psi}=ck\epsilon(y),
\end{eqnarray}
where $c$ is an arbitrary dimensionless parameter and $\epsilon(y)$, which is $1$ for $y>0$ and $-1$ for $y<0$, is responsible for making the mass term even under the ${\bf Z}_2$ symmetry. 

We work in a unitary gauge $A_5=0$ and decompose $A_M$ and $\Psi$ in the KK modes
\begin{eqnarray}
 \Phi(x^\mu,y)={1 \over \sqrt{2\pi r_c}}\sum_{n=0}^\infty \Phi^{(n)}(x^\mu)f_n(y), \label{modeE}
\end{eqnarray}
where $\Phi=\{A_\mu,e^{-2\sigma}\psi\}$ and the KK modes $f_n(y)$ obey the orthonormal condition
\begin{eqnarray}
 {1 \over 2\pi r_c}\int_{-\pi r_c}^{\pi r_c}dy e^{(2-s)\sigma}f_n(y)f_m(y)=\delta_{nm}, \label{on}
\end{eqnarray}
with $s=2,1$ for $A_\mu, \Psi$. Substituting (\ref{modeE}) with solution of $f_n$ into (\ref{action}) and integrating the $y$ direction, we find the gauge coupling of a gauge boson KK mode $n$ to the zero-mode fermion as
\begin{eqnarray}
 g^{(n)}=g_4{1-2c \over e^{(1-2c)\pi kr_c}-1}
{k \over N_n}
 \left[J_1\left({m_n \over k}e^\sigma \right)+b_1(m_n)Y_1\left({m_n \over k}e^\sigma \right)\right], \label{coupling}
\end{eqnarray}
where $J_1$ and $Y_1$ are the standard Bessel functions of the first and
second kind, $N_n$ is a normalization factor, $b_1(m_n)$ is the
constant, $m_n$ is the mass of the $n$th KK mode and $g_4$ is the
four-dimensional $SU(3)$ gauge coupling, related to the five-dimensional
gauge coupling $g_4=g_5/\sqrt{2\pi r_c}$. More detailed discussion is
given in, for example, ref.~\cite{GhPo}. Note that, because of the
orthonormal condition (\ref{on}), self-interactions of the gauge fields
between different modes are not allowed. It means that the KK gluon only
decays to a pair of a quark and an anti-quark. 

We are interested in a situation where the third generation of quarks couples to the KK gluon strongly, compared to the four-dimensional QCD coupling. Under this setup, we study the process $pp\rightarrow g_{\rm KK}^{(1)}\rightarrow b\bar{b}$, where $g_{\rm KK}^{(1)}$ is the first excitation mode of the KK gluon. 
The coupling between the KK mode and the fermions are given by (\ref{coupling}), which is determined by the bulk mass parameter $c$. 
We consider the following scenarios with various values of couplings:
\begin{eqnarray}
&&{g_{Q_3}^{(1)} \over g_4}={g_{t}^{(1)} \over g_4}={g_{b}^{(1)} \over g_4}=4, \quad {g_{\rm light}^{(1)} \over g_4}=0, \label{couplings1}\\
&&{g_{Q_3}^{(1)} \over g_4}=1, \quad {g_{t}^{(1)} \over g_4}={g_{b}^{(1)} \over g_4}=4, \quad {g_{\rm light}^{(1)} \over g_4}=0, \label{couplings2}\\
&&{g_{Q_3}^{(1)} \over g_4}=1, \quad {g_{t}^{(1)} \over g_4}=4,\quad {g_{b}^{(1)} \over g_4}={g_{\rm light}^{(1)} \over g_4}=0, \label{couplings3}
\end{eqnarray}
where $Q_3$ is the third generation of the left-handed quark, $t,b$ are
the right-handed top and bottom quarks and ``light'' means the quarks of
the first two generations. In (\ref{couplings1}), couplings of all the
quarks of the third generation to the KK gluon is strong while the
coupling between the KK gluon and the light quarks is vanishing. 
The latter 
choice is motivated by the constraint coming from the FCNC and the
electroweak precision measurement.  In (\ref{couplings2}),  the KK gluon
strongly couples to the right-handed quarks only. The coupling to the
left-handed quark is comparable to the QCD coupling $g_4$. This choice
has been studied to analyze the decay of the KK gluon to top and anti-top
quarks \cite{LiRaWa}. In (\ref{couplings3}), the difference from
(\ref{couplings2}) is to take the KK gluon coupling to the right-handed
bottom quark to be zero. It is a choice motivated by the constraint from
the flavor physics and the electroweak precision measurement
\cite{JuWe}. With the above choice parameters, we perform the numerical
analysis of the process $pp\rightarrow g_{\rm KK}^{(1)}\rightarrow
b\bar{b}$.

\sect{Numerical Analysis}

We analyze the possibility of the observation of effects of the KK gluon  
predicted by the presented model in the $b\bar{b}$ final states at the
LHC. We simulated the signal $p p \rightarrow \gkk \rightarrow b
\bar{b}$ and possible background processes, initial/final state
radiations, hadronization and decays using Pythia 8.160
\cite{Pythia6,Pythia8}, the Monte-Carlo generator using leading-order
expressions of matrix-elements. All samples were generated for $pp$
collisions at $\sqrt{s} = 14$ TeV using the CTEQ6L1 parton distribution
functions \cite{CTEQ6L1}. The factorization scale $Q_F$ for the
$2\rightarrow 1$ processes ({\it e.g.},  $p p \rightarrow \gkk$) was chosen to be equal to the invariant mass of the final particle, for the $2\rightarrow 2$ processes to be equal to the smaller of the transverse masses of the two outgoing particles. The renormalization scale $Q_R$ for the $2\rightarrow 1$ processes was chosen to be equal to the invariant mass of the final particle, for the $2\rightarrow 2$ processes to be equal to the square root of the product of transverse masses of the two outgoing particles. 
To save time in the massive simulations of $ab\rightarrow cd$ processes, the phase space cuts $\hat{p}_{T}(c) > 50$~GeV, $\hat{p}_{T}(d) > 50$~GeV, $M_{c,d} > 400$~GeV on the transverse momentum of final particles in their center-of-mass system and on their invariant mass were applied.

For the simulation of the effects of a detector, we used Delphes 1.9 \cite{Delphes}, a framework for a fast simulation of a generic collider experiment. The fast simulation of the detector includes a tracking system, a magnetic field of a solenoidal magnet affecting tracks of charged particles, calorimeters and a muon system. The reconstructed kinematical values are smeared according to the settings of the detector simulation. For the jets reconstruction, Delphes uses the FastJet tool \cite{FastJet1,FastJet2} with several implemented jet algorithms. In our simulations, we used the data file with standard settings for the ATLAS detector, provided by the tool. We used the $k_T$ algorithm \cite{kT_algorithm} with a cone radius parameter $R = 0.7$. The $b$-tagging efficiency is assumed to be 40\%, independently on a transverse momentum and a pseudorapidity of a jet. A fake rate of a $b$-tagging algorithm is assumed to be 10\% for $c$-jets and 1\% for light- and gluon-jets. These settings for $b$-tagging are standard for the ATLAS detector in the Delphes 1.9 tool. No trigger inefficiencies are included in this analysis.

\begin{figure}[htb]
\begin{center}
\begin{eqnarray*}
 \begin{array}{cc}
  \epsfxsize=8cm
  \epsfbox{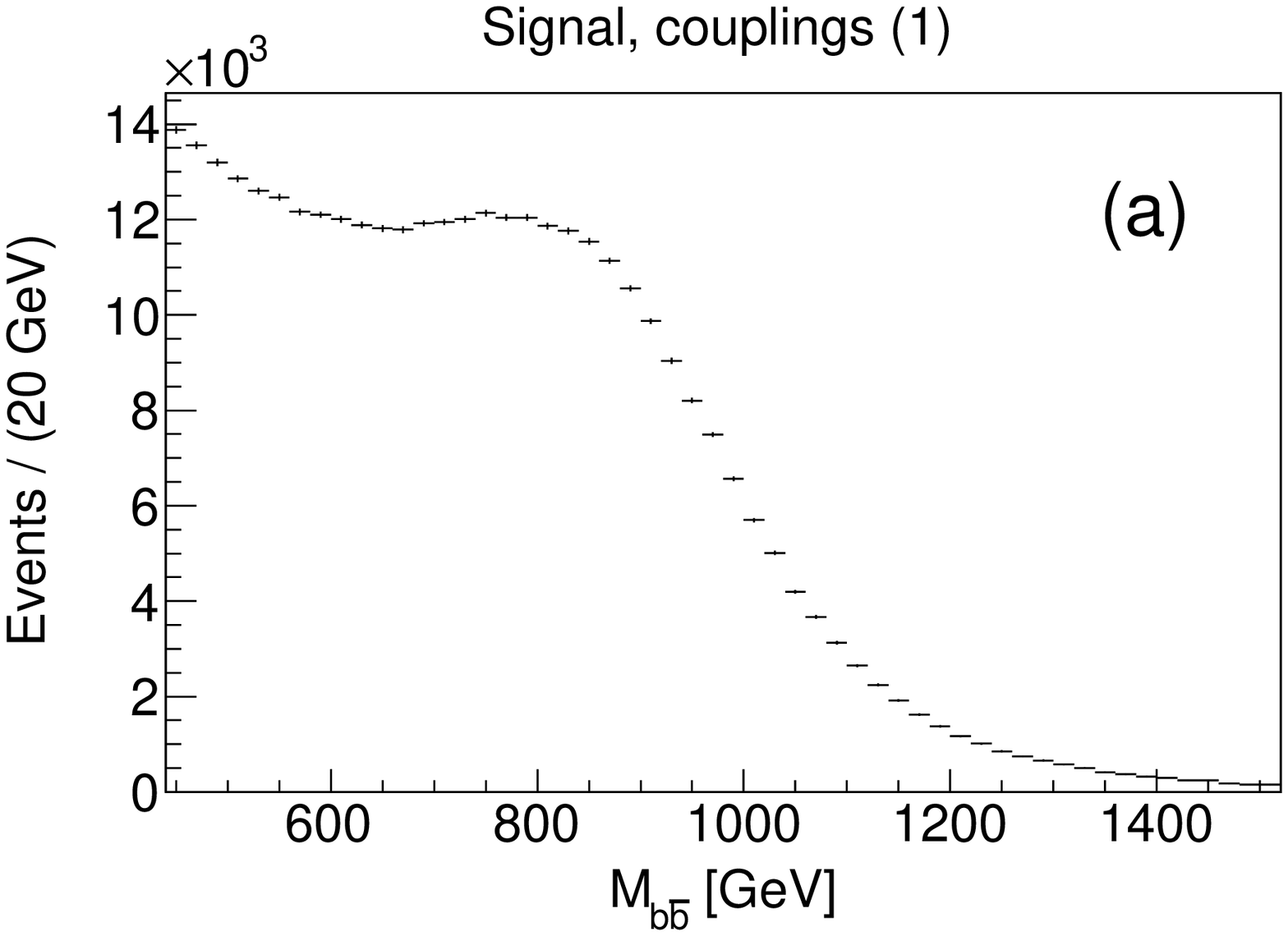}
  &
  \epsfxsize=8cm
  \epsfbox{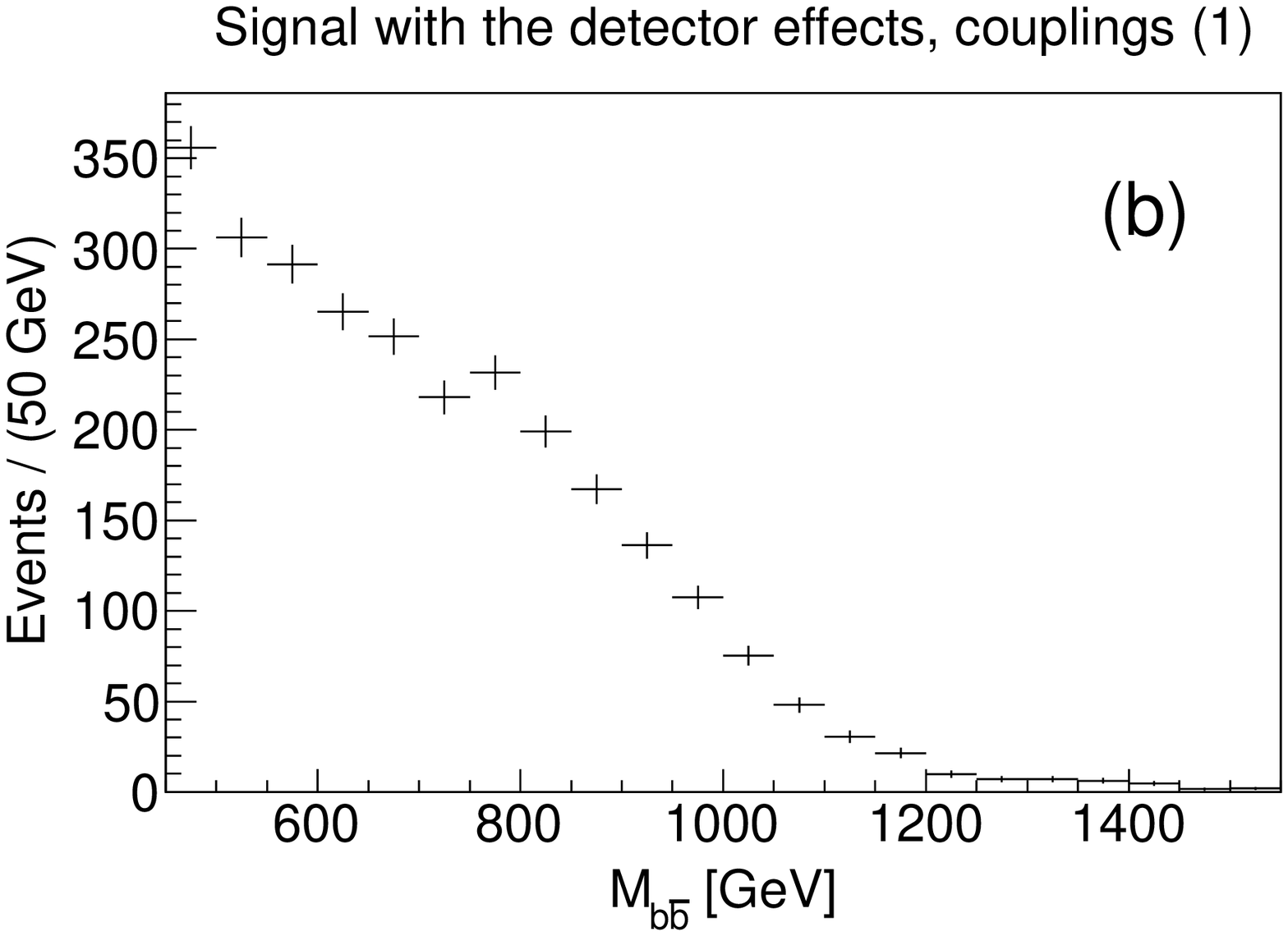} \\
  \epsfxsize=8cm
  \epsfbox{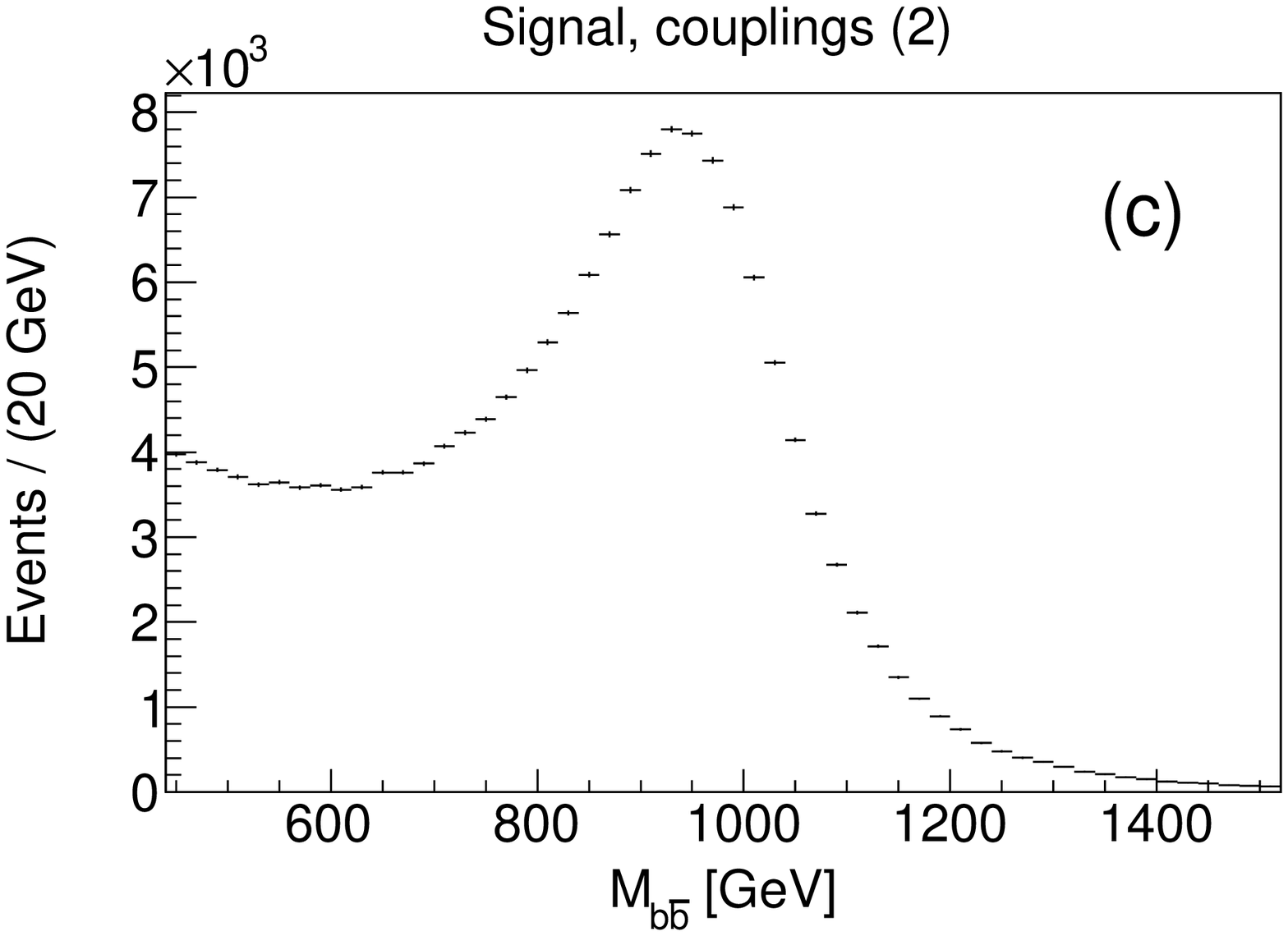}
  &
  \epsfxsize=8cm
  \epsfbox{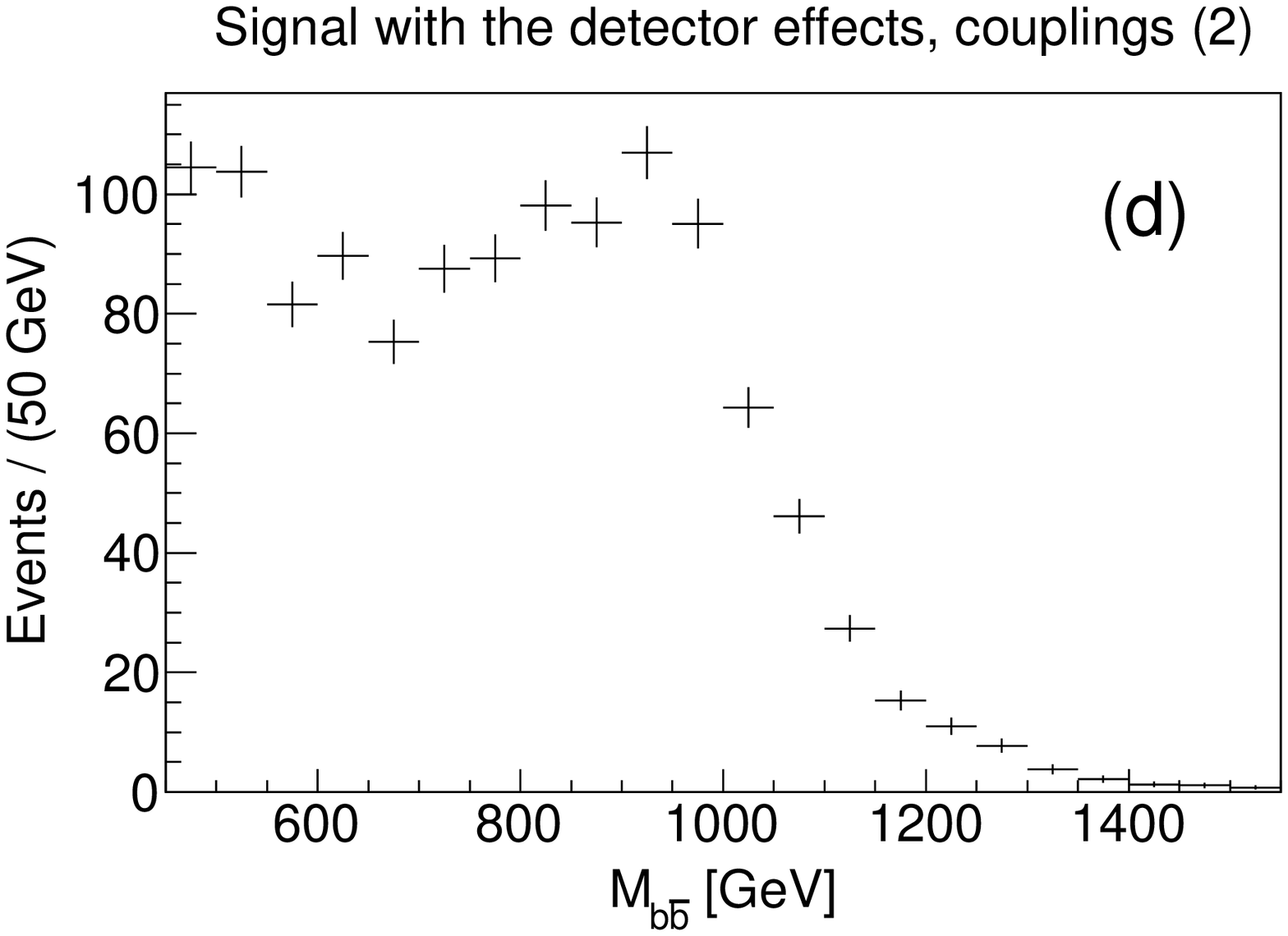} \\
  \epsfxsize=8cm
  \epsfbox{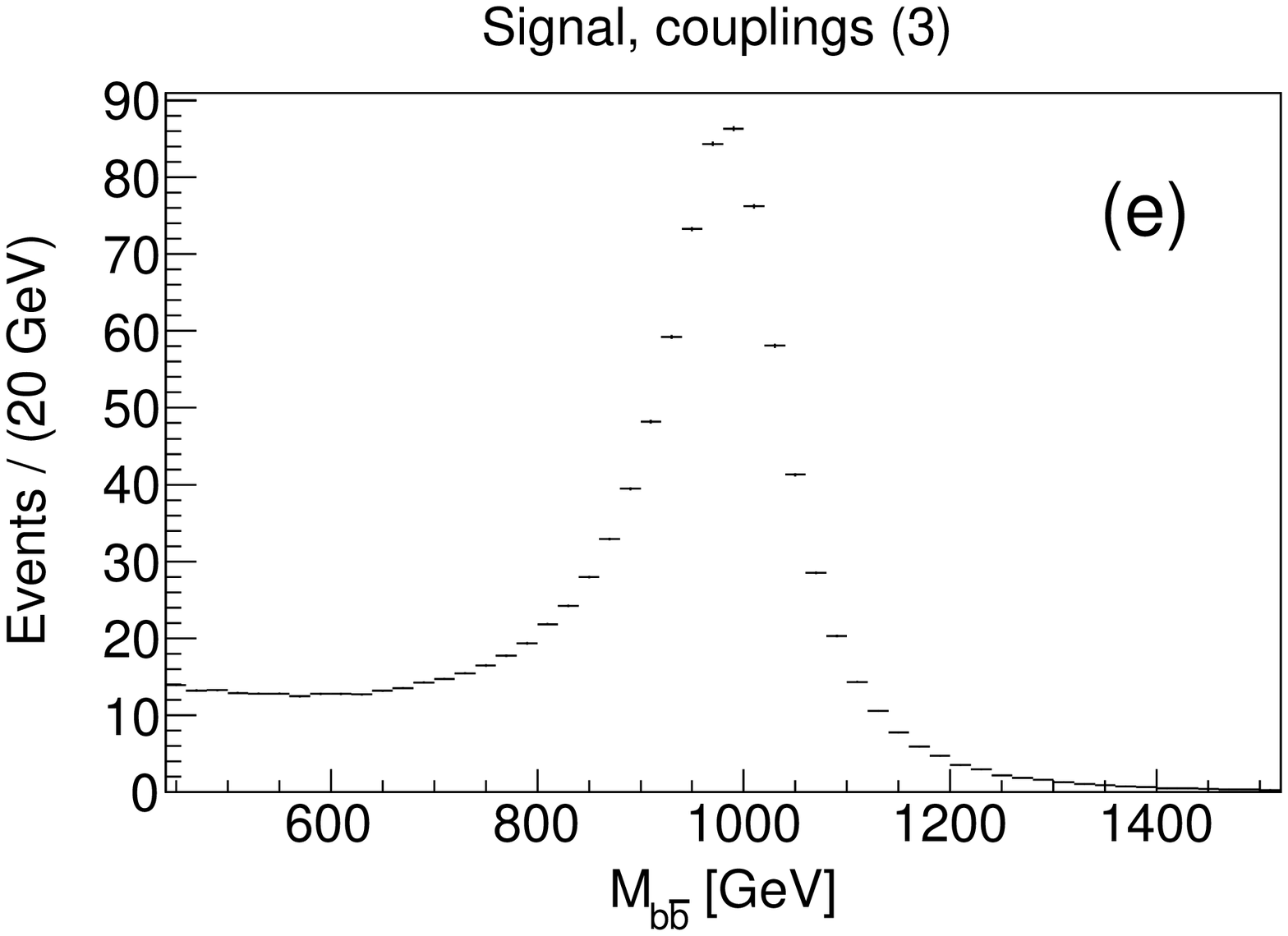}
  & 
  \epsfxsize=8cm
  \epsfbox{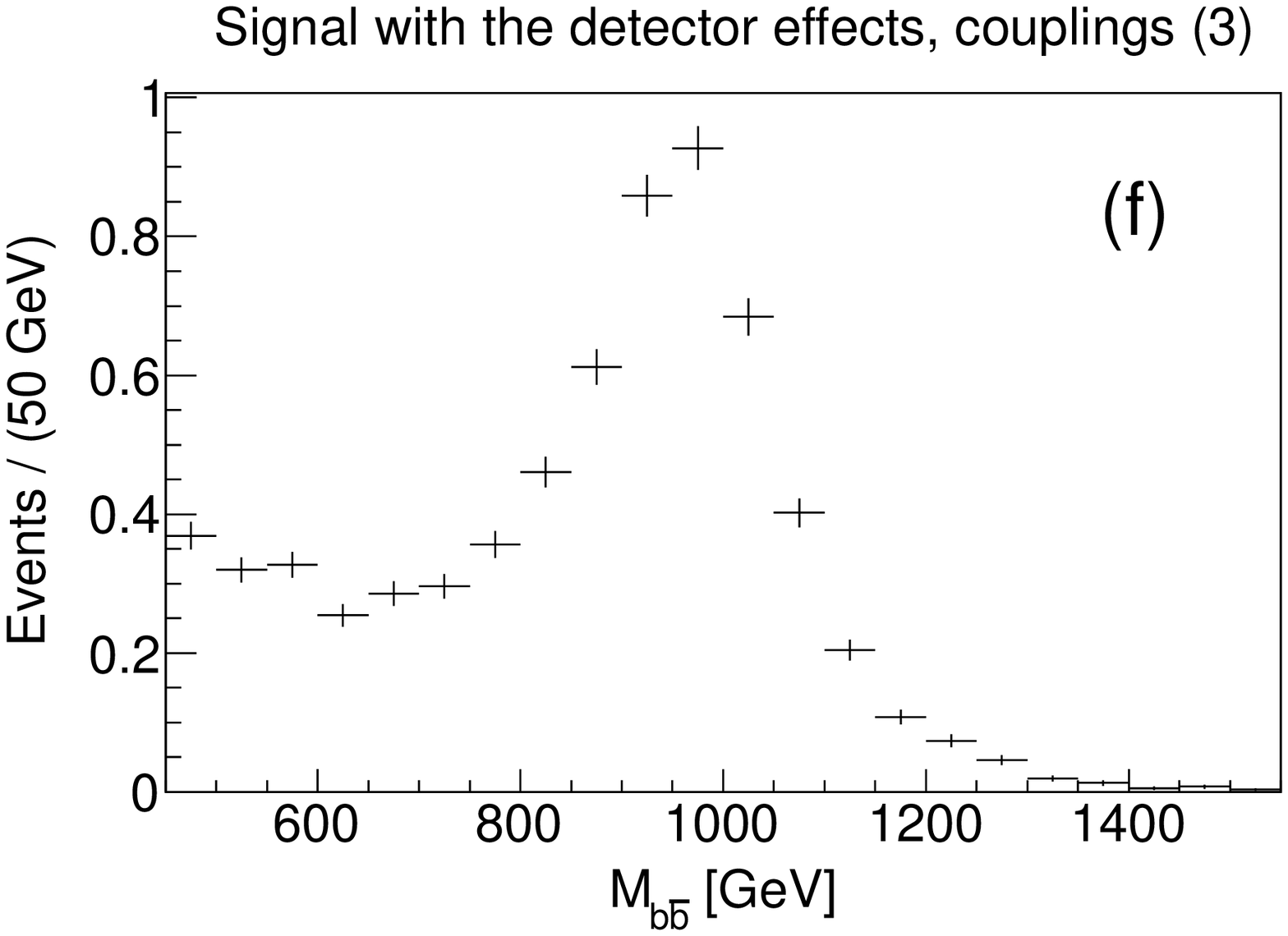} \\
 \end{array}
\end{eqnarray*}
\caption{
The invariant mass distribution of the $b\bar{b}$ pairs for the signal 
 process $pp \rightarrow \gkk \rightarrow b\bar{b}$ without 
 ((a), (c), and (e)) and with ((b), (d), and (f)) the simulated effects 
 of the ATLAS detector and the selection criteria (with 
 $M_{b\bar{b}}^{min}$ = 450 GeV). $M_{\gkk} = 1$~TeV was assumed and 
 three scenarios with couplings (\ref{couplings1})--(\ref{couplings3}) 
 were studied (marked as (1), (2), and (3), in the figure). The number 
 of events in the histogram is scaled to the integrated luminosity of 10 
 fb$^{-1}$ for $pp$ collisions at $\sqrt{s}=14$ TeV.  
}
  \label{signal}
\end{center}
\end{figure}

\begin{figure}[tb]
\begin{center}
  \epsfxsize=12cm
  \epsfbox{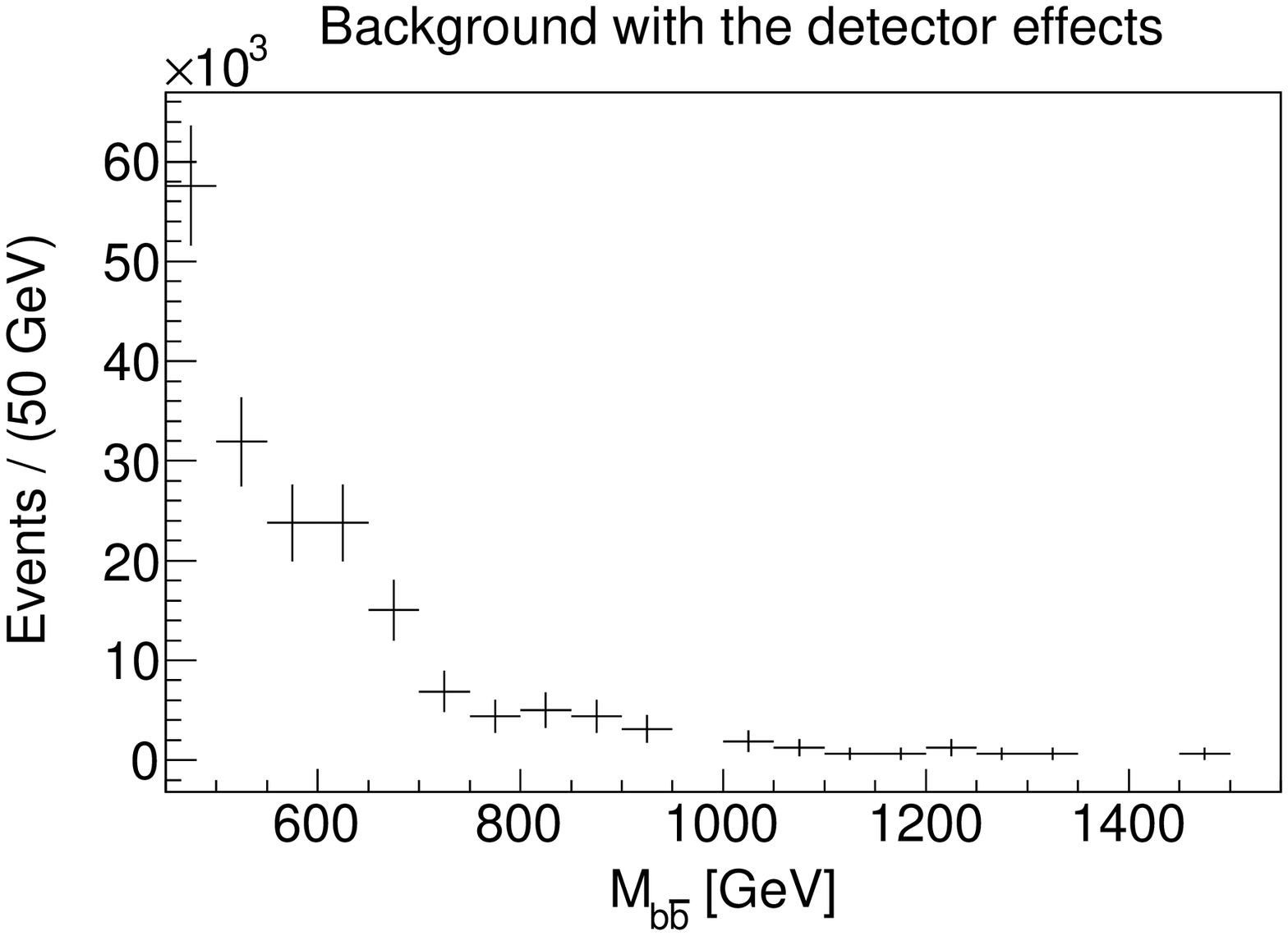} \\
\caption{
The invariant mass distribution of the detected $b$-jets pairs for the background processes with the simulated effects of the ATLAS detector and the selection criteria (with $M_{b\bar{b}}^{min}$ = 450 GeV). The number of events in the histogram is scaled to the integrated luminosity of 10 fb$^{-1}$ for $pp$ collisions at $\sqrt{s}=14$ TeV. 
}
\label{background}
\end{center}
\end{figure}

We assume this set of event selection criteria:
\begin{enumerate}
  \item The event must have exactly 2 $b$-tagged jets with the transverse momentum $p_T > 100$ GeV, the pseudorapidity $|\eta| < 2.5$ and invariant mass $M_{b\bar{b}} > M_{b\bar{b}}^{min}$.
  \item The event must have no other jet with $p_T > 20$ GeV, $|\eta| < 4.9$.
  \item The event must have no electron or muon with $p_T > 10$ GeV, $|\eta| < 2.5$.
  \item The reconstructed transverse missing energy of the event $E_{T}^{miss} < 50$ GeV.
\end{enumerate}
The criterion no. 1 for sufficiently high $M_{b\bar{b}}^{min}$
effectively suppresses the QCD background processes ({\it e.g.},  $gg$ and $q\bar{q}$, $q \in \{u,d,s,c,b\}$, production). The criterion no. 2 suppresses a top-antitop pair production in both the QCD and the RS model, with the subsequent decay of the top quarks to jets. The criteria no. 3 and 4 effectively suppress other decay channels of a top quark decay. 

 We simulated the signal process for couplings
 (\ref{couplings1})--(\ref{couplings3}) and for the masses of the KK gluon
 between 1~TeV and 1.5~TeV. For the analysis, samples of $10^6$ signal
 events were used. Assuming the integrated luminosity of 10 fb$^{-1}$
 per year (during a low luminosity LHC run), it corresponds to the data
 collected during the period with the length from 2.6 up to almost 200
 years, depending on the $\gkk$ mass and couplings. In the 
 Fig. \ref{signal}, distributions of a $b\bar{b}$ invariant mass without
 and with the simulated detector effects and the selection criteria for
 $M_{\gkk} = 1$~TeV and couplings
 (\ref{couplings1})--(\ref{couplings3}) are presented. All plots are
 scaled to the integrated luminosity of 10 fb$^{-1}$.  

For the presented selection criteria, the most important background processes are the QCD production of pairs of (anti)quarks and gluons ($q\bar{q}(gg)\rightarrow gg$, $q'\bar{q'}(gg)\rightarrow q\bar{q}$, $qg\rightarrow qg$, $qq'\rightarrow qq'$; $q,q'\in\{u,d,s,c,b\}$). In the analysis, we used $50\times 10^6$ simulated background events. Due to an extremely high cross section, it corresponds to the data collected only during $1.6\times 10^{-3}$ year (assuming the integrated luminosity of 10 fb$^{-1}$ per year). In the Fig. \ref{background}, the invariant mass distribution of the detected $b$-jets (or objects supposed to be $b$-jets) is plotted. 

\begin{table}[bt]
\centering
\begin{tabular}{c c c c c c c c}
\hline
\multirow{2}{*}{$\frac{g^{(1)}_{\rm light}}{g_4}$} & \multirow{2}{*}{$\frac{g^{(1)}_{Q_3}}{g_4}$} & \multirow{2}{*}{$\frac{g^{(1)}_b}{g_4}$} & \multirow{2}{*}{$\frac{g^{(1)}_t}{g_4}$} & $M_{\gkk}$ & $M_{b\bar{b}}^{min}$ & $S/\sqrt{B}$    & $S/\sqrt{B}$ \\
                              &                          &                       &                       & [TeV]            & [GeV]          & for 10 fb$^{-1}$ & for 100 fb$^{-1}$ \\ 
\hline
\hline
\multirow{5}{*}{0}            & \multirow{5}{*}{4}       & \multirow{5}{*}{4}    & \multirow{5}{*}{4}    & 1.0              & 690            & $7.3 \pm 0.5$    & $24 \pm 2$ \\
                              &                          &                       &                       & 1.1              & 720            & $5.0 \pm 0.4$    & $17 \pm 1$ \\
                              &                          &                       &                       & 1.2              & 720            & $3.4 \pm 0.3$    & $11 \pm 1$ \\ 
                              &                          &                       &                       & 1.3              & 750            & $2.3 \pm 0.2$    & $7.7 \pm 0.7$ \\ 
                              &                          &                       &                       & 1.4              & 940            & $1.7 \pm 0.2$    & $5.7 \pm 0.7$ \\ 
                              &                          &                       &                       & 1.5              & 940            & $1.2 \pm 0.2$    & $4.0 \pm 0.7$ \\ 
\hline
\multirow{5}{*}{0}            & \multirow{5}{*}{1}       & \multirow{5}{*}{4}    & \multirow{5}{*}{4}    & 1.0              & 720            & $4.4 \pm 0.3$    & $15 \pm 1$ \\
                              &                          &                       &                       & 1.1              & 940            & $3.0 \pm 0.4$    & $10 \pm 1$ \\
                              &                          &                       &                       & 1.2              & 940            & $2.0 \pm 0.3$    & $7 \pm 1$ \\ 
                              &                          &                       &                       & 1.3              & 940            & $1.4 \pm 0.2$    & $4.7 \pm 0.7$ \\ 
                              &                          &                       &                       & 1.4              & 1340           & $0.9 \pm 0.5$    & $3 \pm 2$ \\ 
                              &                          &                       &                       & 1.5              & 1340           & $0.9 \pm 0.4$    & $3 \pm 1$ \\ 
\hline
\multirow{1}{*}{0}            & \multirow{1}{*}{1}       & \multirow{1}{*}{0}    & \multirow{1}{*}{4}    & 1.0              & 820            & $0.033 \pm 0.003$ & $0.11 \pm 0.01$ \\
\hline
\end{tabular}
\caption{The statistical significance $S/\sqrt{B}$ of our model for various values of $M_{\gkk}$ and couplings estimated for 10 fb$^{-1}$ and 100 fb$^{-1}$. The presented errors correspond to the statistical errors related to our Monte-Carlo simulations.}
\label{TabSignificance}
\end{table}

As a signature of new physics, we use the number of selected events. For the integrated luminosity of 10~fb$^{-1}$ and 100~fb$^{-1}$, we estimated number of expected observed signal and background events ($S$ and $B$) and the statistical significance $S/\sqrt{B}$. The significance of the deviation from the SM is proportional to the square root of the integrated luminosity. Therefore, it is easy to recompute the results for higher integrated luminosity. We studied effects of variation of $M_{b\bar{b}}^{min}$ on the statistical significance. In the presented results, we use the value of $M_{b\bar{b}}^{min}$, for which the statistical significance $S/\sqrt{B}$ is maximal.

In the Tab.~\ref{TabSignificance}, we present the statistical significance $S/\sqrt{B}$ of our model for various values of
$M_{\gkk}$ and couplings. 
As expected, the deviation from the SM is strongly dependent on the coupling of a right-handed $b$ quark to a KK gluon. 
For the first set of couplings (\ref{couplings1}), the effects of KK gluons could be observable with the significance of $5\sigma$ for the mass of a KK gluon up to 1.1 TeV and the integrated luminosity of 10 fb$^{-1}$ or for the mass of a KK gluon up to 1.4 TeV and the integrated luminosity of 100 fb$^{-1}$. For the second set of couplings (\ref{couplings2}), the effects of KK gluons could be observable with the significance of $5\sigma$ for the mass of a KK gluon up to 1.2 TeV and the integrated luminosity of 100 fb$^{-1}$. Due to the extremely low cross-section of the signal process, for the third set of couplings (\ref{couplings3}) the effects of KK gluons are unobservable.

\sect{Summary}

We studied the possibility of observation of effects of the first excitation of a KK gluon, predicted by the extension of the RS model. In our work, we aimed on the final states with two $b$ jets. We prepared appropriate Monte-Carlo simulations of the signal and background processes for the $pp$ collisions with the energy $\sqrt{s} = 14$~TeV at the LHC, simulated the effects of the ATLAS detector and the selection criteria. As a signature of new physics, we used the number of selected events. We studied three scenarios (\ref{couplings1})--(\ref{couplings3}) with various couplings of a KK gluon to $b$ and $t$ quarks. We estimated the significance $S/\sqrt{B}$ of our model. For the integrated luminosity of 100~fb$^{-1}$, the effects of a KK gluon will be observable with significance $5\sigma$ in the scenario (\ref{couplings1}) with strong coupling of $b$ and $t$ quarks to a KK gluon for the mass of a KK gluon up to 1.4~TeV. In the scenario (\ref{couplings2}), when a KK gluon strongly couples to right-handed $b$ and $t$ quarks only, the effects of new physics will be observable for the mass of a KK gluon up to 1.2~TeV. Even from the integrated luminosity of 10~fb$^{-1}$, the deviation from the
SM could be observable with the significance of several sigmas for the mass of a KK gluon up to 1.5~TeV and scenarios (\ref{couplings1}) and (\ref{couplings2}). The effects of a KK gluon in the scenario (\ref{couplings3}) with vanishing coupling of a KK gluon to a right-handed $b$ quark will not be observable.  

\vspace{1cm}

\noindent
{\bf Acknowledgements:}\\
The work of M.A. and K.S. is supported in part by the Research Program
MSM6840770029, by the project of International Cooperation ATLAS-CERN 
and by the project LH11106 of the Ministry of Education, Youth and Sports 
of the Czech Republic. 
The work of G.C.C is supported in part by Grants-in-Aid for Scientific 
Research from the Ministry of
Education, Culture, Sports, Science and Technology (No.24104502) and 
from the Japan Society for the Promotion of Science (No.21244036).

\small{

\end{document}